\documentclass[twocolumn]{aastex63}

\usepackage{amsmath}
\usepackage{amssymb}
\usepackage{graphicx}
\usepackage{xcolor}
\usepackage{multirow}

\newcommand{\Om}{\Omega_\mathrm{m}}
\newcommand{\Ob}{\Omega_\mathrm{b}}
\newcommand{\s}[1]{\sigma_{#1}}
\newcommand{\ee}[1]{\times10^{#1}}
\newcommand{\Msun}{\mathrm{M}_\odot}
\newcommand{\unit}[1]{\,\mathrm{#1}}
\newcommand{\hMpc}{h^{-1}\mathrm{Mpc}}

\shorttitle{Impact of Baryons on Cosmological Inference from WL}
\shortauthors{Lu et~al.}

\begin{document}

\title{The Impact of Baryons on Cosmological Inference from Weak Lensing Statistics}

\correspondingauthor{Tianhuan Lu}
\email{tl2854@columbia.edu}

\author[0000-0003-1040-2639]{Tianhuan Lu}
\affiliation{Department of Astronomy, Columbia University, New York, NY 10027, USA}

\author[0000-0003-3633-5403]{Zolt\'an Haiman}
\affiliation{Department of Astronomy, Columbia University, New York, NY 10027, USA}

\begin{abstract}

As weak lensing surveys are becoming deeper and cover larger areas, information will be available on small angular scales down to the arcmin level. To extract this extra information, accurate modelling of baryonic effects is necessary. In this work, we adopt a baryonic correction model, which includes gas both bound inside and ejected from dark matter (DM) haloes, a central galaxy, and changes in the DM profile induced by baryons. We use this model to incorporate baryons into a large suite of DM-only $N$-body simulations, covering a grid of 75 cosmologies in the $\Om-\s8$ parameter space.  We investigate how baryons affect Gaussian and non-Gaussian weak lensing statistics and the cosmological parameter inferences from these statistics. Our results show that marginalizing over baryonic parameters degrades the constraints in $\Om-\s8$ space by a factor of $2-4$ compared to those with baryonic parameters fixed.  We investigate the contribution of each baryonic component to this degradation, and find that the distance to which gas is ejected (from AGN feedback) has the largest impact due to its degeneracy with cosmological parameters. External constraints on this parameter, either from other datasets or from a better theoretical understanding of AGN feedback, can significantly mitigate the impact of baryons in an HSC-like survey.
\end{abstract}


\section{Introduction}
\label{sec:introduction}

Weak gravitational lensing (hereafter weak lensing, or WL) describes the bending of light rays coming from distant objects by inhomogeneities in the foreground matter distribution. Since the matter distribution depends on the underlying cosmological model, WL was proposed and has been proven to be a powerful tool in inferring cosmological parameters, such as $\Om$ and $\s8$ in a $\Lambda$CDM universe \citep[see, e.g.][for reviews]{bartelmann2001, refregier2003b, hoekstra2008,kilbinger2015}. Recent studies using data from WL surveys have yielded especially strong constraints on a certain combination of cosmological parameters (e.g. $S_8\equiv\s8(\Om/0.3)^{0.5}$) and are approaching percent-level precision \citep{joudaki2016, kohlinger2017, hikage2019, hamana2020}. 

One of the trends in the existing and planned WL survey programs is the increase in depth---the number density of galaxies has increased from $\sim10\unit{arcmin^{-2}}$ in the past to $\sim20\unit{arcmin^{-2}}$ at present, and up to $50\unit{arcmin^{-2}}$ is expected in the future \citep{heymans2012, dejong2013, des2005, aihara2018, laureijs2011, spergel2015, ivezic2019}. Although this will allow accessing more information theoretically on small scales (down to a few arcmin), making use of this information is challenging since many systematic errors are more prominent on small scales if not treated accurately.  These effects include uncertainties in baryonic physics \citep{zhan2004, white2004, jing2006} and intrinsic alignments of galaxy shape orientations \citep{schneider2010, sifon2015}.

In this paper, we focus on investigating how the baryonic physics affect WL statistics --- specifically the (convergence) power spectrum and WL peak counts (as an example of a small-scale non-Gaussian statistic). Previous studies have incorporated baryonic physics into WL observables using two broadly different approaches. The first is to utilize hydrodynamical simulations that assume certain treatments of baryonic physics and predict modifications to WL statistics. For example, \citet{hikage2019} take the fitting formula based on the active galactic nucleus (AGN) feedback model in the OverWhelmingly Large Simulations project \citep{schaye2010}, considering AGN feedback having the largest effect, and adjust the power spectrum. \citet{osato2021} have recently studied the baryonic effects using ray-traced convergence maps based on the IllustrisTNG hydrodynamics simulations, which include several baryonic effects \citep{nelson2019}. The second approach is to modify the shape of the dark matter (DM) haloes in a DM-only $N$-body simulations via so-called ``baryonic corrections''. An early attempt following this approach studying baryonic effects on cosmological parameter estimates include \citet{yang2013}. This work analysed how varying the concentration parameters of Navarro-Frenk-White (NFW) haloes affects the shear power spectrum and the lensing peak counts. More recently, \citet{schneider2019} compared power spectra and shear correlation functions adopting more sophisticated baryonic corrections, and showed that they can accurately mimic the observables inferred from hydrodynamical simulations.

In this paper, we take the latter approach, employing a state-of-the-art baryonic correction model (hereafter BCM) proposed by \citet[][a modification of an earlier version by \citealt{schneider2015}]{arico2020}, to answer two questions:
\begin{itemize}
\item How much does baryonic physics affect the lensing power spectrum and peak counts?
\item If we infer baryonic model parameters simultaneously with cosmological parameters, how well can they be constrained?
\end{itemize}
The BCM offers a parametrized way to characterize how baryonic physics alter the shape of each halo on the basis of a DM-only $N$-body simulation, where each parameter has a physically motivated interpretation. Compared to hydrodynamical simulations, the BCM is much less computationally expensive, while retaining flexibility to model various poorly constrained aspects of baryonic physics. \cite{arico2020} have shown that the matter power spectrum residuals between various hydrodynamical simulations and their best-fitting BCMs are under 1\%. 

We organize this paper as follows. In \S~\ref{sec:methods}, we introduce our simulation suite, how we derive statistics from simulations, and the BCM. In \S~\ref{sec:results}, we show the effects of baryonic physics and the posterior distributions of cosmological and baryonic parameters from WL. In \S~\ref{sec:discussion}, we then discuss how factors such as different noise levels and restrictions of power spectrum measurements to different angular scales affect parameter inferences, as well as the implications of our findings. Finally, we summarize our main conclusions in \S~\ref{sec:conclusions}.

\section{Methods}
\label{sec:methods}

\subsection{$N$-body simulations}

\begin{figure}[!ht]
\centering
\includegraphics[width=7.5cm]{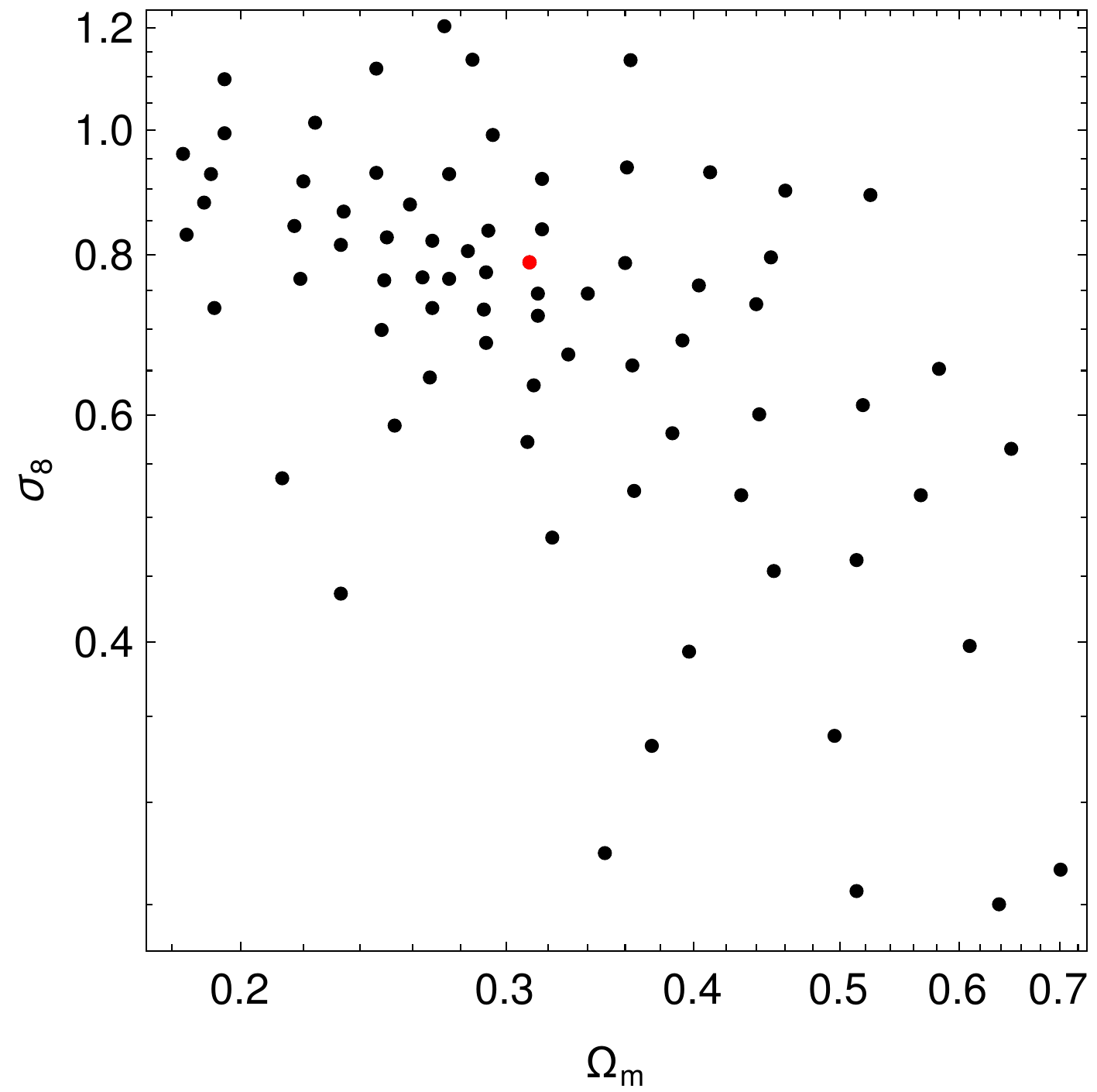}
\caption{The cosmologies of the 75 $N$-body simulations in the suite that we use in this study. The red dot represents the fiducial cosmology ($\Om=0.311, \s8=0.789$).}
\label{fig:cosmologies}
\end{figure}

Our simulation suite consists of DM-only $N$-body simulations of $\Lambda$CDM cosmologies with 75 different combinations of $\Om$ (the matter density) and $\s8$ (the magnitude of matter fluctuation). Our parameter grid is illustrated in Figure~\ref{fig:cosmologies}.

The rest of the cosmological parameters are fixed as follows: Hubble constant $H_0=72\unit{km\,s^{-1}\,{Mpc}^{-1}}$, baryon density $\Ob=0.046$, scalar spectral index $n_\mathrm{s}=0.96$, effective number of relativistic degrees of freedom $n_\mathrm{eff}=3.04$, and neutrino masses $m_\nu=0$.

The initial conditions are computed using \textsc{CAMB} \citep{lewis2000}, and the simulations are run with the $N$-body code \textsc{Gadget-2} \citep{springel2005}. The size of each simulation run is $240\,\hMpc$ (comoving) with $512^3$ DM particles, each of which has a mass of $\approx 10^{10}\,\Msun$. In the fiducial cosmology ($\Om=0.311, \s8=0.789$), the size of the simulation box corresponds to a field of view of $6\times6\unit{{deg}^2}$ at redshift $z=1$.

\subsection{Ray-tracing}

We follow the procedures described by \citet{petri2016b} to generate the WL convergence ($\kappa$) maps from $N$-body simulations using the multiple lens plane algorithm \citep{jain2000, hilbert2009}. We briefly introduce the main steps in this procedure below and refer the reader to the above papers for more details.

In this study, we assume that all lensed galaxies are at $z=1$. We take snapshots of each $N$-body simulation at a series of redshifts between $z=0$ and $z=1$, such that the difference in the comoving distance between the redshifts of adjacent snapshots is $80\,\hMpc$. For each snapshot in the series, we cut a slab of width $80\,\hMpc$ along a random axis ($x$, $y$ or $z$), and we generate a \textit{density plane} by calculating the column density of a slab on a $4096\times4096$ grid followed by random shifts and rotations (the latter in multiples of $90^\circ$). To obtain a \textit{potential plane}, we solve the two-dimensional Poisson equation on each density plane for the gravitational potential field. Finally, we follow light rays from $z=0$ to $z=1$ using the multiple lens plane algorithm on each series of potential planes following \citet{petri2016b} within a fixed field of view of $3.5\times3.5\unit{{deg}^2}$ (consistent with previous works on this simulation suite which ray-traced to $z=2$). The result of this procedure is a $2048\times2048$ pixelized converge map. Through cutting slabs and shifting and rotating density planes randomly, we can repeat this process to generate multiple realizations of the convergence map for each simulation, which can be considered statistically independent \citep{petri2016a}.

To imitate the convergence maps calculated from real data, we add shape noise to our simulated convergence maps by changing the value of each pixel by a random number, drawn from a Gaussian distribution:
\begin{equation}
\left\{\mu=0, \sigma=\frac{\sigma_\epsilon}{\sqrt{2 n_\mathrm{gal} A_\mathrm{pix}}}\right\}, 
\end{equation}
where $\sigma_\epsilon=0.4$ is the typical mean intrinsic ellipticity of galaxies, $n_\mathrm{gal}$ the surface density of lensed galaxies controlling the level of noise, and $A_\mathrm{pix}$ the area of each pixel. We also smooth the maps with a $1\unit{arcmin} (\sim10\unit{pixels})$ Gaussian filter to remove artefacts caused by pixelation and to facilitate peak counting statistics. 

\subsection{Modification by baryons}

\begin{table*}[!ht]
\centering
\begin{tabular}{cccc}
\hline
Parameter & Fiducial value & Prior bound \\
\hline
$M_\mathrm{c}$ & $3.3\ee{13}h^{-1}\Msun$  & $\left[5.9\ee{11},4.4\ee{15}\right]h^{-1}\Msun$ \\
$M_{1,0}$      & $8.63\ee{11}h^{-1}\Msun$ & $\left[9.3\ee{10},1.1\ee{13}\right]h^{-1}\Msun$ \\
$\eta$         & 0.54                     & $\left[0.12,2.7\right]$ \\
$\beta$        & 0.12                     & $\left[0.026,3.8\right]$ \\
\hline
\end{tabular}
\caption{The fiducial values and prior bounds of the baryonic parameters}
\label{tab:baryon-priors}
\end{table*}

The BCM assumes that each halo can be divided into four components---DM, stars, bound gas, and ejected gas. The radial profiles and relative weights of these components are controlled by four baryonic parameters: $M_\mathrm{c}$, the halo mass for retaining half of the total gas; $M_{1,0}$, the characteristic halo mass for a galaxy mass fraction of 0.023 at $z=0$; $\eta$, the maximum distance to which the gas content of a halo is ejected (in units of a characteristic escape distance that is roughly 5 virial radii); and $\beta$, the slope of the gas fraction against the halo mass.  We choose the fiducial values for the baryonic parameters following \cite{arico2020}, and adopt wide prior bounds so that they cover sufficient variations in the matter power spectrum (see Table~\ref{tab:baryon-priors}). 

To incorporate baryonic effects into the $N$-body simulations, we replace DM haloes with analytical profiles that include baryons. Specifically, we remove all particles which are considered to belong to DM haloes, and paint the pixelated image of column density of the spherically symmetric haloes (with baryons) on the corresponding density planes. 

\begin{figure}[!ht]
\centering
\includegraphics[width=8cm]{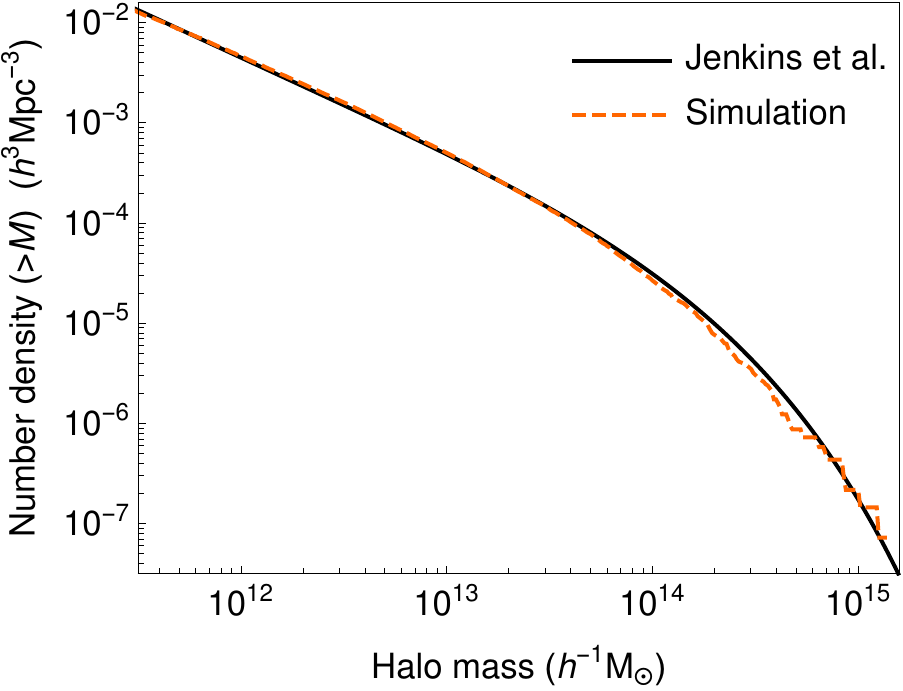}
\caption{The cumulative halo mass function at $z=0$ from the fiducial simulation compared to that given by \citet{jenkins2001}.}
\label{fig:halomass}
\end{figure}

First, we build halo catalogues for every snapshot of each $N$-body simulation with \textsc{rockstar} \citep{behroozi2012}, a halo finder based on adaptive hierarchical refinement of friends-of-friends (FoF) groups, see Figure~\ref{fig:halomass} for one of the halo mass functions. We set the FoF linking length to the default value of 0.28. \textsc{rockstar} assigns a list of DM particles to each halo or subhalo, where each particle belongs to no more than one list, and we catalog all haloes with total particle mass greater than $10^{12}\,\Msun$. Second, for each halo, we remove its particles from the slab if the distance between the particle and the halo centre is smaller than the virial radius ($r_{200}$). We refer to the total mass of this region inside $r_{200}$ as the halo mass $M$ (note that some of the particles assigned by \textsc{rockstar} to a halo are outside $r_{200}$; for our purposes those particles remain outside the halo). We then fit the radial distribution of the removed particles to the NFW profile with mass $M$, and obtain the best-fitting concentration parameter $c$ of each halo from a maximum likelihood estimate. Third, given a specific choice of baryonic parameters, we calculate the column density of each halo (analytical profile with BCM) on the density plane grid at their locations, and they are added to the density plane. 

In this work, we generate one realization of the convergence map for each choice of baryonic parameters for each cosmology. The baryonic parameters are chosen according to a Sobol sequence \citep{sobol1967} of 160 elements in the 4-dimensional prior hypercube. Sobol sequences have a property of low-discrepancy, i.e. evenly distributed across the space, so they perform better in polynomial fitting and interpolation compared to random points. Compared to Latin hypercube sampling, Sobol sequences have the advantage of being extendable, in case we need more elements in the parameter space to improve accuracy. In addition, we generate 128 realizations at the fiducial baryonic parameters to estimate the covariance matrix for each cosmology.

We note that there are, on average, $\sim30,000$ haloes in an $N$-body snapshot, and  $\gtrsim10^{10}$ 2D halo images with BCM need to be calculated given our setup  (75 cosmologies $\times$ $(160+128)$ realizations $\times$ 30 snapshots $\times$ $30,000$ haloes). It is too computationally expensive to calculate every halo following the method proposed by \cite{arico2020} and \citet{schneider2015}, since that technique involves displacing individual N-body particles.  To reduce the computational cost, we compute the column densities of the four components in BCM separately:
\begin{align}
\nonumber \Sigma_\mathrm{halo}(\theta_\mathrm{b}, z, M, r_\mathrm{200},  c)&=M\,w_\mathrm{dm}(\theta_\mathrm{b}, z)\,\Sigma_\mathrm{dm}(r_\mathrm{200}, c) \\
\nonumber &+M\,w_\mathrm{star}(\theta_\mathrm{b}, z)\,\Sigma_\mathrm{star}(r_\mathrm{200}, c) \\
\nonumber &+M\,w_\mathrm{bg}(\theta_\mathrm{b}, z)\,\Sigma_\mathrm{bg}(r_\mathrm{200}, c) \\
&+M\,w_\mathrm{eg}(\theta_\mathrm{b}, z)\,\Sigma_\mathrm{eg}(\eta, r_\mathrm{200}),
\end{align}
where $\theta_\mathrm{b}$ denotes all baryonic parameters. $\Sigma_\mathrm{dm}$ and $\Sigma_\mathrm{bg}$ are the profiles of the DM and the bounded gas respectively, almost all the masses of which reside inside $r_\mathrm{200}$. $\Sigma_\mathrm{star}$ is the profile of stellar mass \citep{mohammed2014} concentrated at the centre of the halo; it is also referred to as the central galaxy component. $\Sigma_\mathrm{eg}$ is the profile of the ejected gas modelled by particles from AGN feedback following the Maxwell–Boltzmann distribution, the radius of which relative to the escape distance $\approx 5\,r_\mathrm{200}$ is determined by $\eta$. Here, the profiles of the baryonic components contain their contributions to the relaxation of DM $\Sigma_b^\mathrm{(rdm)} (b=\mathrm{\{star, bg, eg\}})$:
\begin{equation}
\Sigma_b^\mathrm{(rdm)}=\frac{1}{\delta}\left(\Sigma_\mathrm{rdm}(\delta\Sigma_b)-\Sigma_\mathrm{dm}\right),
\end{equation}
where $\Sigma_\mathrm{rdm}(\delta\Sigma_b)$ is the profile of relaxed DM with $\delta$ of the halo mass replaced by the baryonic profile $\Sigma_b$. We note that this simplification only retains the linear dependency of  relaxed DM on the baryonic components.

We calculate three of the four surface density profiles $\Sigma_{\{\mathrm{dm, star, bg}\}}$ on a grid of
\begin{align}
\log_{10}\left(\frac{r_{200}}{l_\mathrm{pix}}\right)&=0.12,0.16,\cdots,1.8, \\
\log_{10}c&=0.1,0.2,\cdots,1.8, \\
\log_{10}\left(\frac{r_{200}}{l_\mathrm{pix}}\right)&=0.12,0.16,\cdots,1.8, \\
l_\mathrm{pix}&=\frac{240\,\hMpc}{4096},
\end{align}
with the total mass of each profile being unity, and additionally the contribution on relaxed DM by ejected gas $\Sigma_\mathrm{eg}^\mathrm{(rdm)}$ on
\begin{equation}
\log_{10}\eta=-1.0,-0.9,\cdots,1.0.
\end{equation}
These profiles are then rasterized into images with their centers shifted by 0.0, 0.25, 0.5, and 0.75 pixels along both axis. Whenever we need the image of an actual halo with BCM, we find the images of its components with the closest $M$ and $c$ from the pre-calculated gallery, and weight them by $w_{\{\mathrm{dm, star, bg, eg}\}}$ according to BCM. 

\subsection{Statistics and parameter inference}

In this study, we employ two statistics on the convergence maps---power spectrum and peak counts. The power spectrum of a convergence map is defined as the Fourier transform of the two-point correlation function. We calculate the power spectra by first performing Fourier transforms on the convergence maps, and then obtain the values in 18 equally-spaced logarithmic bins of the angular Fourier mode $\ell$ within $100<\ell<12,000$. Peak counts are defined as the number of peaks per unit solid angle at difference convergence ($\kappa$) values, where a peak refers to a pixel that have a higher $\kappa$ than all of its eight neighboring pixels. We divide $-0.03<\kappa<0.15$ into 18 equally spaced bins and count the number of peaks with $\kappa$ values within each of these bins.

We estimate the posterior distribution with Bayes' theorem on six parameters $\theta$ with log-uniform priors: $0.2<\Om<0.6$, $0.4<\s8<1.1$ and the four baryonic parameters shown in Table~\ref{tab:baryon-priors}. The likelihood function is given by
\begin{gather}
p(\mathbf{y}_0|\theta)\propto\frac{1}{\sqrt{\det\mathbf{C}}}\exp\left(-\frac{1}{2}\Delta\mathbf{y}^\mathrm{T}\widehat{\mathbf{C}^{-1}}\Delta\mathbf{y}\right), \label{eqn: delta-y}\\
\Delta \mathbf{y}=\mathbf{y}_0-\mathbf{y}(\theta), \\
\widehat{\mathbf{C}^{-1}}=\frac{N-d-2}{N-1}\mathbf{C}^{-1}, \\
\mathbf{C}=\left(\frac{A_\mathrm{survey}}{3.5\times3.5\unit{deg^2}}\right)^{-1}\mathbf{C}_\mathrm{sim}(\Om,\s8),
\end{gather}
where $\mathbf{y}(\theta)$ denotes the statistics at parameters $\theta$, $\mathbf{y}_0$ the statistics at the fiducial parameters, $N$ the number of realizations, $d$ the number of observables, $A_\mathrm{survey}$ the area of the supposed survey (e.g. $A_\mathrm{survey}=1,500\unit{deg^2}$ for a survey such as the one by Hyper Suprime-Cam or HSC; \citealt{aihara2018}), and $\mathbf{C}_\mathrm{sim}(\Om,\s8)$ the covariance matrix measured from the simulated convergence maps. In general, the covariance matrix depends on the choice of both cosmological and baryonic parameters, but we ignore its dependence on baryonic parameters here because: 1) the dependence of the covariances on baryonic parameters is small (see \S~\ref{sec:cov-dependence} below) and 2) a single realization for each choice of baryonic parameters will not yield meaningful estimations. The factor $(N-d-2)/(N-1)$ before the covariances makes $\widehat{\mathbf{C}^{-1}}$ and unbiased estimation of the precision matrix \citep{hartlap2007}.

Both statistics (as well as their combination) and their covariances are interpolated across the parameter space (on logarithmic scales) by a fifth-order polynomial to cover arbitrary points over the full range of models. We find that to fit 12,000 points (75 cosmologies $\times$ 160 baryon models), a fifth-order 6-variate polynomial (462 parameters)
%can you explain the 462?
captures the variation of the statistics on all parameters with good accuracy without overfitting. Then, we sample the posterior distribution with a Monte-Carlo Markov chain using differential evolution for $10^6$ steps, which is sufficient for the chains to converge. 

\section{Results}
\label{sec:results}

\subsection{Effects of halo replacement}
\label{sec:results-replacement} 

\begin{figure*}[!ht]
\centering
\includegraphics[width=17cm]{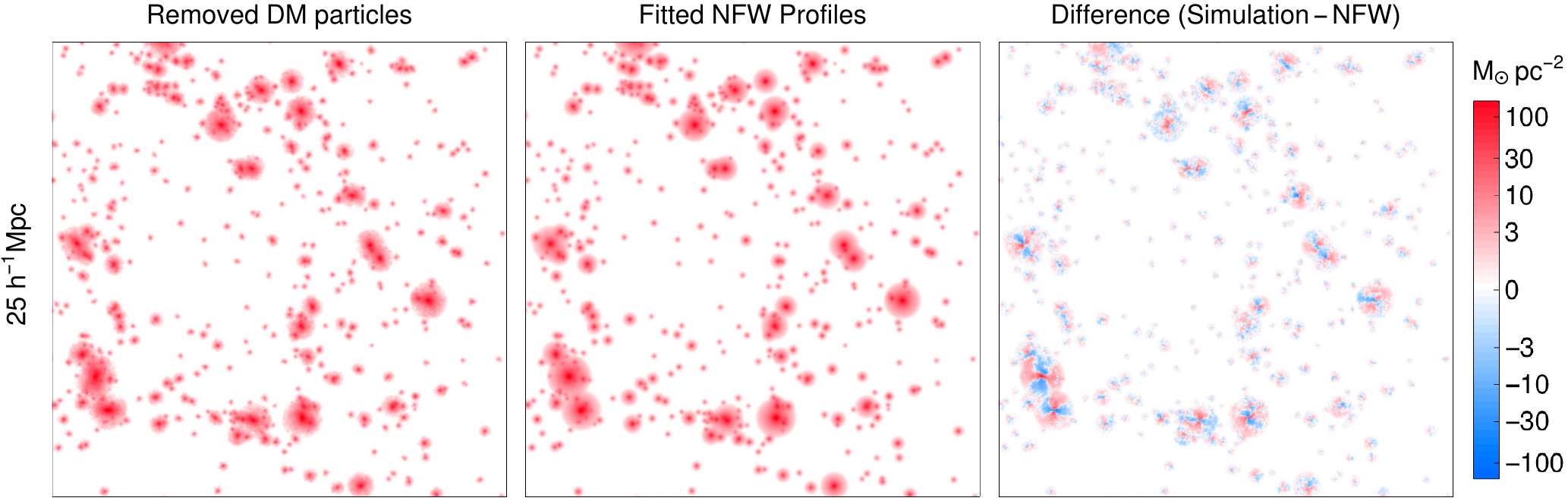}
\caption{The column density of the removed DM particles, the NFW profile images, and their differences in a $(25\,\hMpc)^2$ region of a slab at $z=0$. The maps are smoothed at a $100\unit{kpc}$ scale for better visibility.}
\label{fig:halo-replacement}
\end{figure*}

\begin{figure}[!ht]
\centering
\includegraphics[width=8cm]{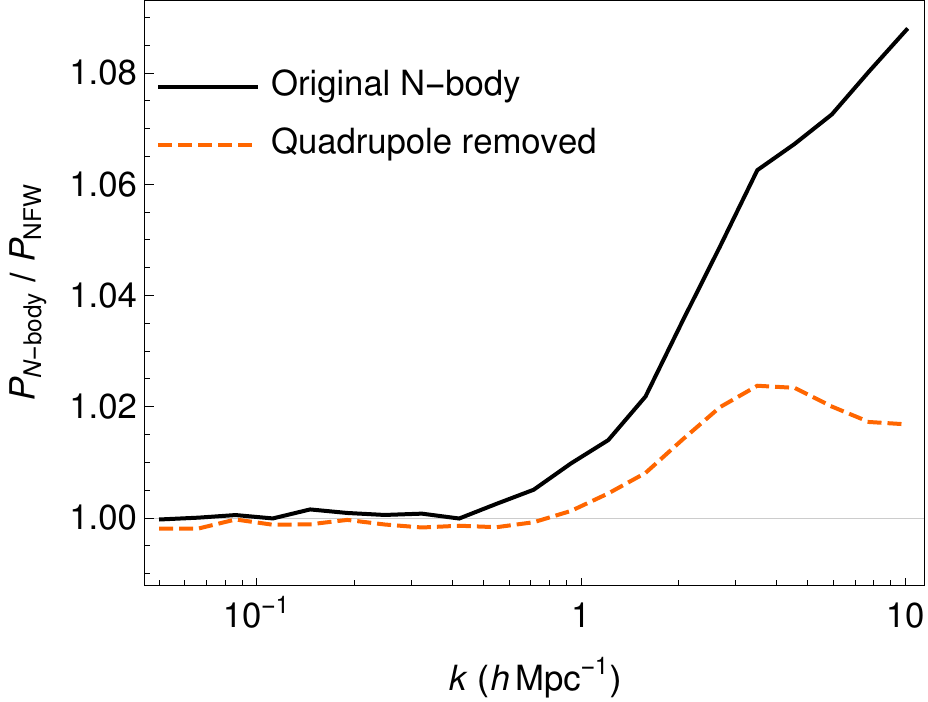}
\caption{The average column density power spectra from the full $N$-body simulations (the original in black, and with halo quadrupoles removed in orange) compared to the power spectrum $P_{\rm NFW}$ after halo replacement in the fiducial cosmology at $z=0$.}

\label{fig:halo-replacement-ps}
\end{figure}

\begin{figure}[!ht]
\centering
\includegraphics[width=8cm]{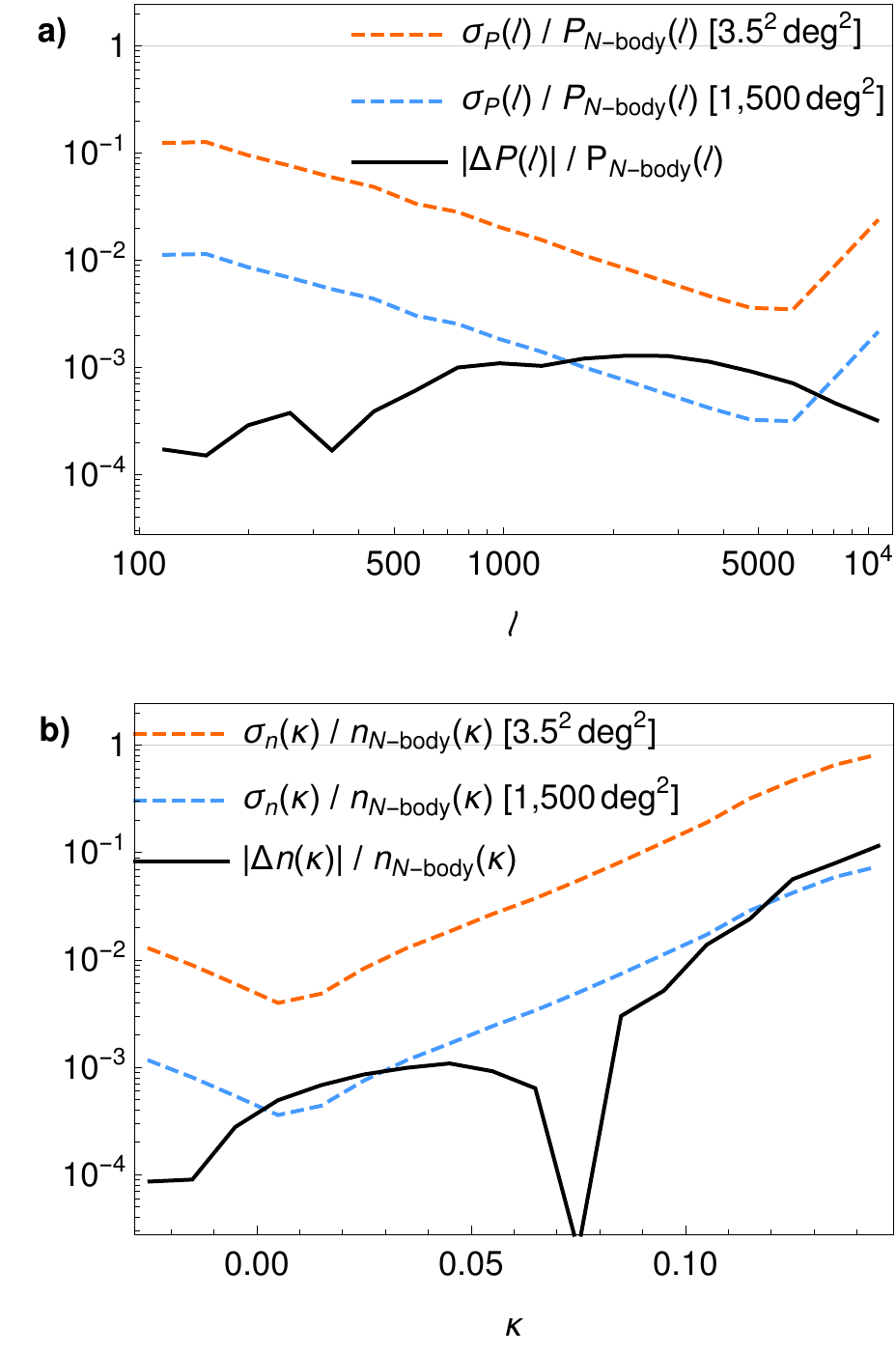}
\caption{The residuals caused by halo replacement (solid lines) compared to the uncertainties of the observables (dashed lines) in the two statistics, \textbf{a)} power spectrum and \textbf{b)} peak counts.} 
\label{fig:halo-replacement-pspc}
\end{figure}

The process of halo replacement speeds up the calculations of baryon models, but it will introduce systematics since the haloes in the $N$-body simulations does not match NFW profile exactly. In this section, we will study by how much this mismatch affects the statistics.

Figure~\ref{fig:halo-replacement} shows an example of how replacing haloes with NFW profiles changes the column density in a simulation slab, and Figure~\ref{fig:halo-replacement-ps} shows the change in column density power spectrum for the fiducial cosmology at $z=0$. We find that halo replacement lowers the column density power spectrum by $\sim7\%$ mostly at very small scales ($k\gtrsim3\,h\unit{Mpc^{-1}}$). It can be seen from the residual map shown in right panel of Figure~\ref{fig:halo-replacement} that the most noticeable difference for each halo between the $N$-body particles and the NFW profile is the quadrupole component. This means that projecting haloes to a 2-dimensional map leaves their quadrupole components stronger than the other multipole moments and they are apparent when comparing to isotropic NFW profiles. To investigate the importance of this quadrupole residual, we average the original column density map and its duplicate but with the halo particles rotated around the projected halo centre by $90^\circ$, and we repeat the comparison of the column density power spectrum (see Figure~\ref{fig:halo-replacement-ps}). As the discrepancy shrinks to $\sim2\%$, we conclude that the quadrupole residuals are the primary causes of the difference in the small-scale power spectrum introduced by the adoption of spherical NFW haloes.

Although quadrupole residuals introduce systematics in general, we did not attempt to correct them in this work because they have a limited impact on the lensing power spectrum and peak counts as all maps are affected in the same way (but we note that they may be a concern for other statistics or when our method is applied on real data). In Figure~\ref{fig:halo-replacement-pspc}, we compare the residuals caused by halo replacement to the uncertainty of the observables from 64 realizations of the fiducial cosmology with a noise level of $n_\mathrm{gal}=20\unit{arcmin^{-2}}$. We find that for both statistics, the residuals are mostly less than 10\% of the uncertainties of the observables in a single convergence map, but they are projected to be comparable to the uncertainties in a $1,500\unit{deg^2}$ survey. As a result, a correction for these quadrupolar asymmetries will need to be incorporated into the spherical halo-based baryon corrections, when real data is fit in a large ($\gtrsim 1500~{\rm deg^2}$) future survey.

\subsection{Impact of baryons on statistics}
\label{sec:results-baryon} 

\begin{figure}[!ht]
\centering
\includegraphics[width=8cm]{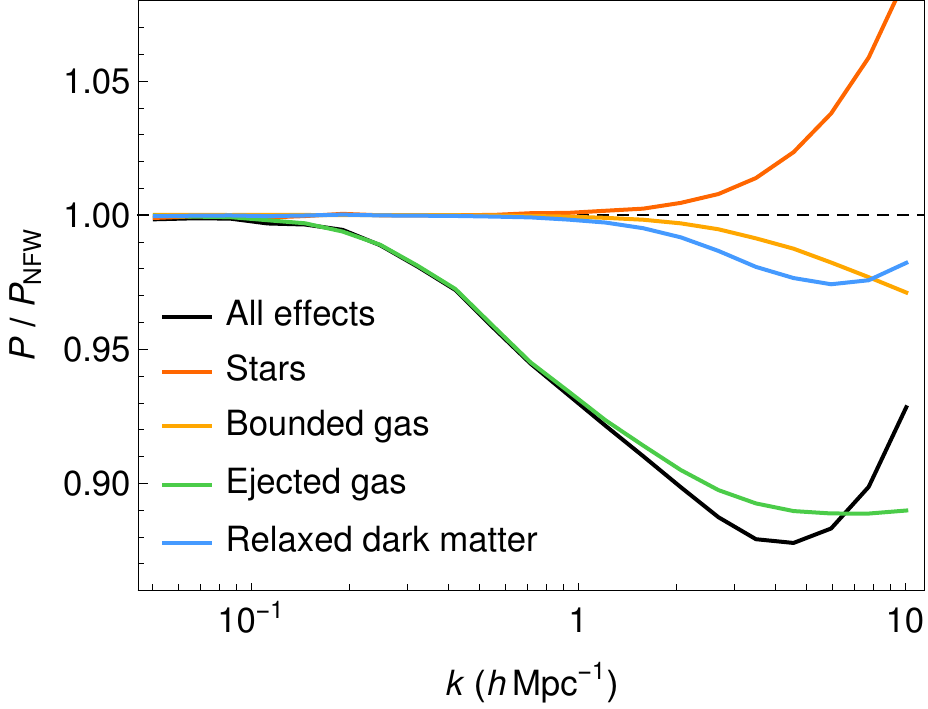}
\caption{The column density power spectra with each of the four baryon components and with all components at $z=0$, taking fiducial values. They are compared to the power spectrum from replacing haloes with NFW profiles without baryons.}
\label{fig:baryon-components}
\end{figure}

\begin{figure*}[!ht]
\centering
\includegraphics[width=14.5cm]{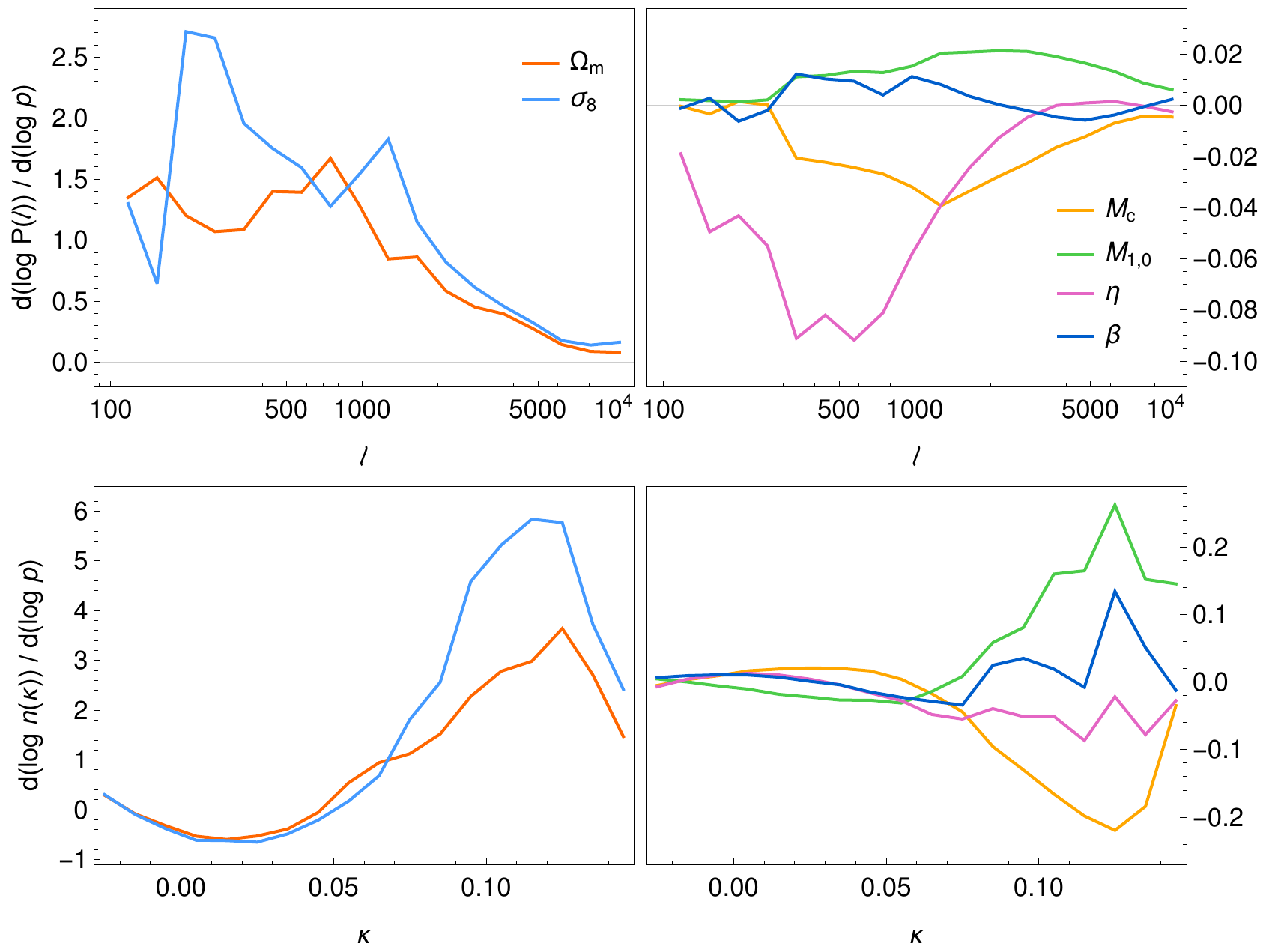}
\caption{The responses of the power spectrum (top panels) and peak counts (bottom panels) to variations in the cosmological (left) and baryonic (right) parameters at their fiducial values in an HSC-like survey area (1,500 deg$^2$) and galaxy density (20 amin$^{-2}$. }
\label{fig:parameter-response}
\end{figure*}

In Figure~\ref{fig:baryon-components}, we show the column density power spectra with each of the four baryon components included.  The trends in this figure are very similar to Figure~3 in \citet{arico2020} except that they were showing 3-dimensional matter power spectra. A key difference in implementing BCM between our work and \citet{arico2020} is that we replace halo particles by the images of analytical profiles with baryons, while \citet{arico2020} moves individual DM particles to match the effects caused by baryonic physics. We note that our method has the advantage of being more accurate in representing baryonic effects when there are only a small number of particles in the haloes, but it has the disadvantage of not retaining the original shapes of the haloes and replaces them with spherically symmetric profiles (see \S~\ref{sec:results-replacement} above), which can induce biases in statistics. Regardless of these differences in baryonification, our method produces matter power spectra of baryon components that are very similar to \citet{arico2020}.    Figure~\ref{fig:baryon-components} shows that among these components, the ejected gas (AGN feedback) has by far the largest affect on the power spectrum, suppressing it by $7\%$ at $k=1h\unit{Mpc^{-1}}$.

Figure~\ref{fig:parameter-response} shows how much the statistics will change if we tweak each cosmological and baryonic parameter, where we choose an HSC-like scenario: survey area $A_\mathrm{survey}=1,500\unit{deg^2}$ and galaxy density $n_\mathrm{gal}=20\unit{arcmin^{-2}}$. We find that the sensitivity of both statistics to baryonic parameters are much smaller than to cosmological parameters by a factor of $\sim10-100$, which means the capability of power spectrum and peak counts in constraining baryonic parameters is correspondingly weaker. We also find that the responses show degeneracies between certain cosmological and baryonic parameters. Most prominently, the response of both the power spectrum and the peak counts to $\eta$ is qualitatively similar to its response to variations in $\Om$, in the opposite direction.

\subsection{Parameter inference with baryons}
\label{sec:results-inference}

\begin{figure*}[!ht]
\centering
\includegraphics[width=17.5cm]{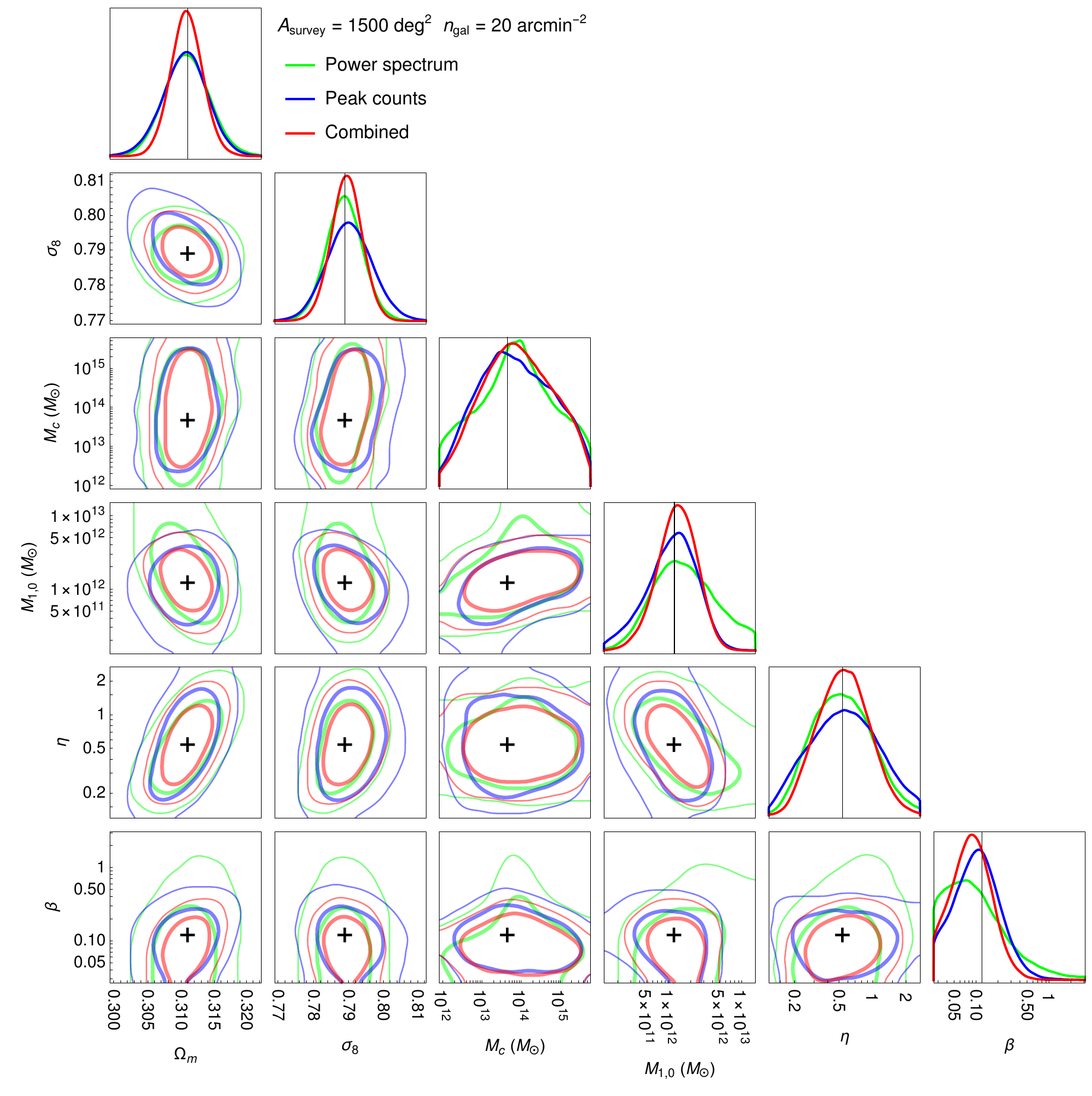}
\caption{The posteriors of cosmological and baryonic parameters in an HSC-like survey area and galaxy density. The thick and thin contours show the $1\sigma$ (68\%) and $2\sigma$ (95\%) credible region respectively, and the black crosses and lines show the fiducial values of the parameters.}
\label{fig:cornerplot-rt-1500-20}
\end{figure*}

\begin{table*}[!ht]
\centering
\begin{tabular}{c|ccc|ccc}
\hline
\multirow{2}{*}{Methods} &  \multicolumn{3}{c|}{$n_\mathrm{gal}=20\unit{arcmin^{-2}}$}  &  \multicolumn{3}{c}{$n_\mathrm{gal}=50\unit{arcmin^{-2}}$} \\
& $S_\mathrm{full}/10^{-5}$ & $S_\mathrm{fid}/10^{-5}$ & $S_\mathrm{full}/S_\mathrm{fid}$ & $S_\mathrm{full}/10^{-5}$ & $S_\mathrm{fid}/10^{-5}$ & $S_\mathrm{full}/S_\mathrm{fid}$ \\ \hline
Power spectrum    & 12.5 & 4.5 & 2.8 &  8.3 & 2.8 & 3.0 \\
Peak counts       & 14.0 & 4.2 & 3.4 & 13.5 & 2.8 & 4.7 \\
Combined          &  7.3 & 2.4 & 3.0 &  4.2 & 1.5 & 2.9 \\
PS ($\ell<2,000$) & 13.9 & 5.9 & 2.4 & 11.0 & 4.3 & 2.5 \\ \hline
\end{tabular}
\caption{The area of the 68\% posterior distributions in $\Om-\s8$ space in a survey with an area of $1500\unit{deg^2}$ and two different galaxy densities, marginalized over baryonic parameters ($S_\mathrm{full}$) or conditioned on fiducial baryonic parameters ($S_\mathrm{fid}$). }
\label{tab:posterior-area}
\end{table*}

\begin{figure}[!ht]
\centering
\includegraphics[width=8cm]{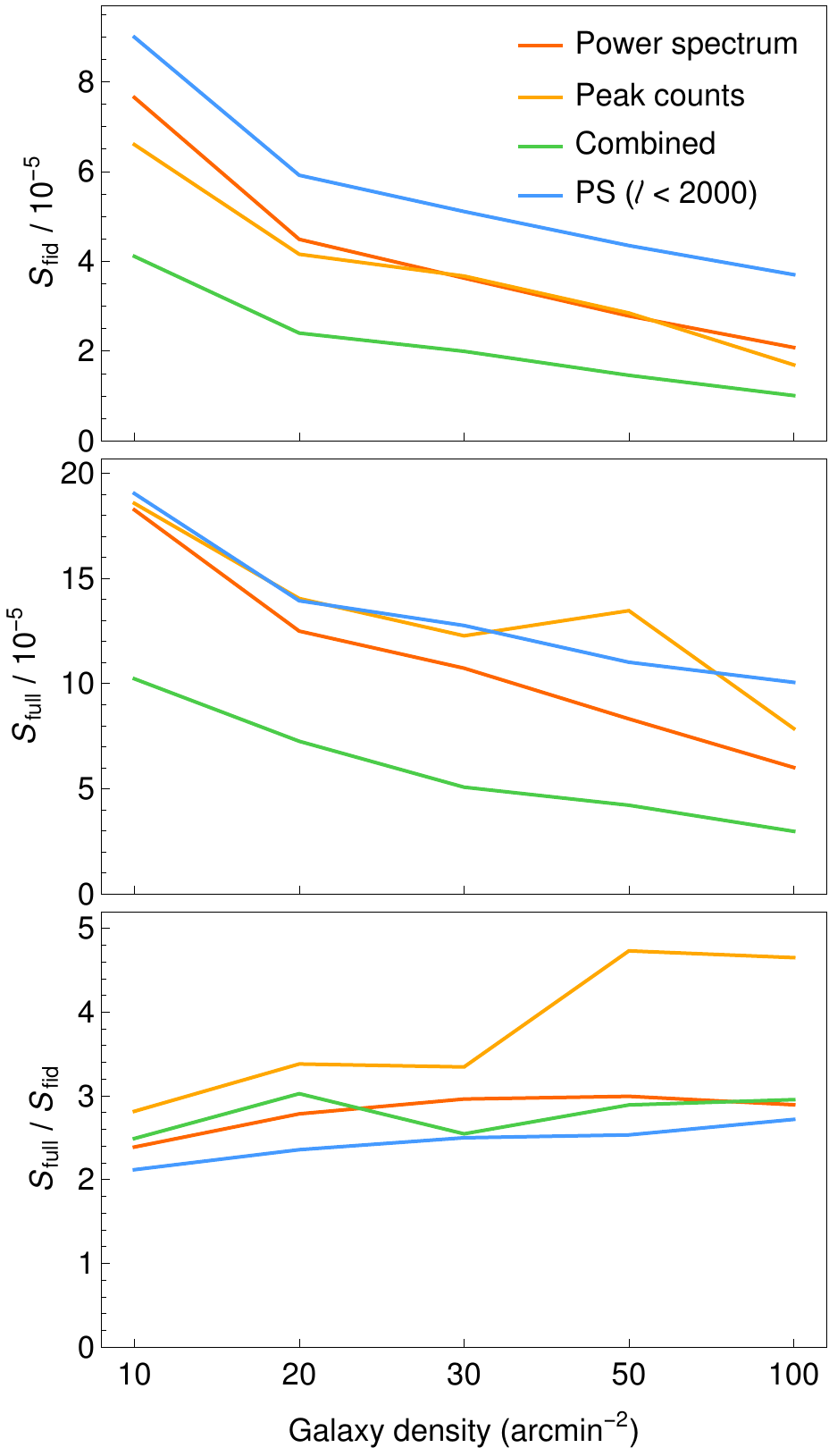}
\caption{The area of the posterior distributions using the same setting as Table~\ref{tab:posterior-area} but including higher galaxy densities. }
\label{fig:posterior-area}
\end{figure}

Figure~\ref{fig:cornerplot-rt-1500-20} shows the posteriors of cosmological and baryonic parameters in an HSC-like survey calculated as described in \S~\ref{sec:methods}. To quantify the ability in constraining cosmological parameters, we marginalize over baryonic parameters and calculate the area of the posterior distribution $S_\mathrm{full}$ in $\Om-\s8$ space enclosed by the $1\sigma$ (68\%) credible contour. The areas of the three statistics (see Table~\ref{tab:posterior-area}) are $1.2\ee{-4}$ (power spectrum), $1.4\ee{-4}$ (peak counts), and $0.7\ee{-4}$ (combined).

Among all 2-dimensional marginal distributions of a cosmological and a baryonic parameter, $\eta$ shows the largest correlation with cosmological parameters, which is consistent with the analysis on sensitivities in \S~\ref{sec:results-baryon}. We define another area of the posterior in $\Om-\s8$ space $S_\mathrm{fid}$ by fixing the baryonic parameters to their fiducial values, so that $S_\mathrm{full}/S_\mathrm{fid}$ represents the factor by which the joint constraints on cosmological parameters would degrade if a method were to constrain baryonic parameters at the same time. We find that in an HSC-like survey, the factor of this degradation is 2.8 for the power spectrum, 3.4 for the peak counts, and 3.0 for these two combined (see Figure~\ref{fig:posterior-area}). This suggests that the two statistics have similar degeneracies between cosmological and baryonic parameters, and combining them does not significantly reduce the degradation. This is again consistent with the findings in Section~\ref{sec:results-baryon}, which show similar degeneracies with $\eta$ for both the power spectrum and the peak counts.

\section{Discussion}
\label{sec:discussion}

\subsection{Parameter inference at different noise levels}

In Figure~\ref{fig:posterior-area}, we show the $1\sigma$ area of the posterior in $\Om-\s8$ space at different noise levels marginalized over baryonic parameters or conditioned on fiducial baryonic parameters. We assume a survey with HSC-like area ($1500\unit{deg^2}$) and galaxy densities from $10\unit{arcmin^{-2}}$ to $100\unit{arcmin^{-2}}$. We find that at higher galaxy densities (lower noise levels), the constraints on $\Om$ and $\s8$ generally becomes tighter, meanwhile the degradation due to marginalizing over the baryonic parameters is larger, especially for peak counts. We conclude that the statistics can extract more information from small scales/high peaks but only significantly so when the baryonic physics is specified. 

\subsection{Comparing power spectrum in full range with $\ell<2,000$}

When the lensing power spectrum is used to constrain cosmological parameters, the upper limit of $\ell$ is typically restricted to $\lesssim2,000$ \citep[e.g.][]{kohlinger2016, kohlinger2017, hikage2019}. The motivation is that at even higher $\ell$ (smaller scales), the matter and lensing power spectra might be affected by baryonic physics to a degree that is hard to model accurately. In this section, we explore the improvements in constraining cosmological parameters when the power spectrum at small scales is included. Figure~\ref{fig:posterior-area} compares the results with the power spectrum over the full range ($100<\ell<12,000$) with that restricted to large scales only ($100<\ell<2,000$), where we only take the first 11 bins and the covariance matrices are adjusted accordingly. 

When the baryonic parameters are free, we find that the small-scale power spectrum provides an 11\% reduction in the $1\sigma$ area in an HSC-like survey; when the baryonic parameters are fixed, this reduction increases to 24\%. If the galaxy density is increased from $20\unit{arcmin^{-2}}$ to $50\unit{arcmin^{-2}}$ matching the noise level of the Nancy Grace Roman Space Telescope, the reductions will be further increased to 25\% (baryons free) and 36\% (baryons fixed). As comparison, \citet{fang2007} found ($n_\mathrm{gal}\approx30\unit{arcmin^{-2}}$, no baryons) that by increasing the upper limit of $\ell$ from 1,000 to 3,000, the uncertainty in $\Om$ and $\s8$ are improved by 30\% and 50\% respectively. 

Nominally, a comparison by using the fixed-baryons constraints restricted to large angular scales gives better constraints ($S_{\rm fid}\approx 6$ for at $20\unit{arcmin^{-2}}$; blue curve in Figure~\ref{fig:posterior-area}) than including smaller scales but at the cost of marginalising over the baryons ($S_{\rm full}\approx 12$; red curve). However, we find this conclusion to be unjustified. Contrary to the notion that baryons only affect small scales, we find that restricting the power spectrum to $\ell<2,000$ does not reduce the degradation caused by uncertainties in the baryonic parameters. Table~\ref{tab:posterior-area} shows that $S_\mathrm{full}/S_\mathrm{fid}$ for the large-scale power spectrum equals $2.4$, which is only modestly lower than that of the full-range power spectrum (by $0.4$). This means that taking baryonic physics into consideration is necessary even for the large-scale power spectrum, at least in the context of the BCM models adopted here.

\subsection{Priors on baryonic parameters}

\begin{figure}[!ht]
\centering
\includegraphics[width=8cm]{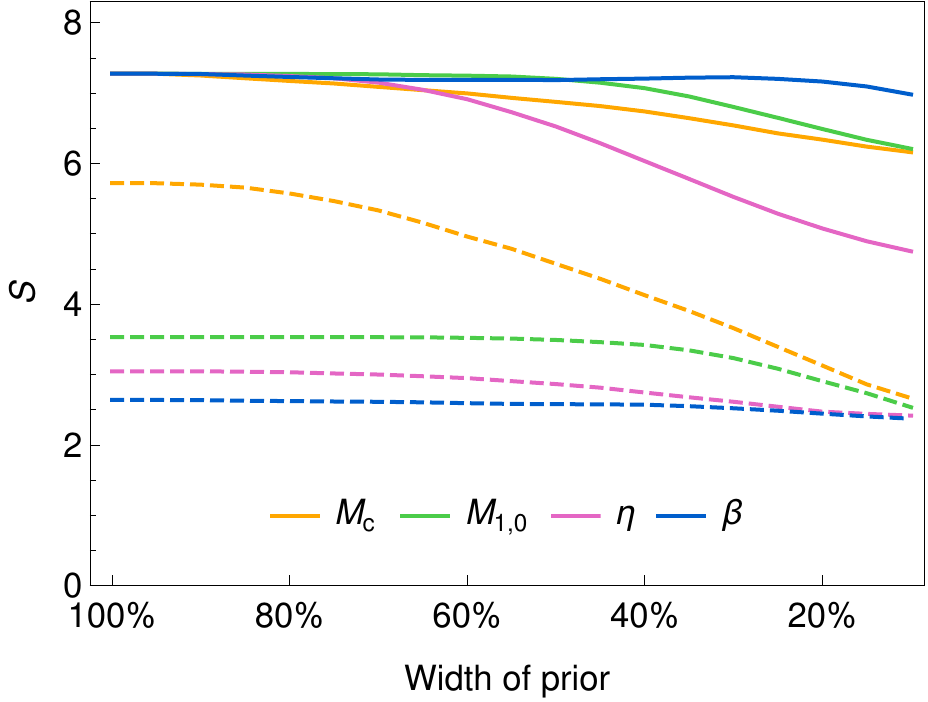}
\caption{$1\sigma$ confidence area in $\Om-\s8$ space for the power spectrum and peak counts combined. The solid lines show one case where the prior of one baryonic parameter is varied and the others are free. The dashed lines show another case, where the prior of one baryonic parameters is varied and the other parameters are fixed at their fiducial values.}
\label{fig:prior-bounds-all}
\end{figure}

\begin{figure}[!ht]
\centering
\includegraphics[width=8cm]{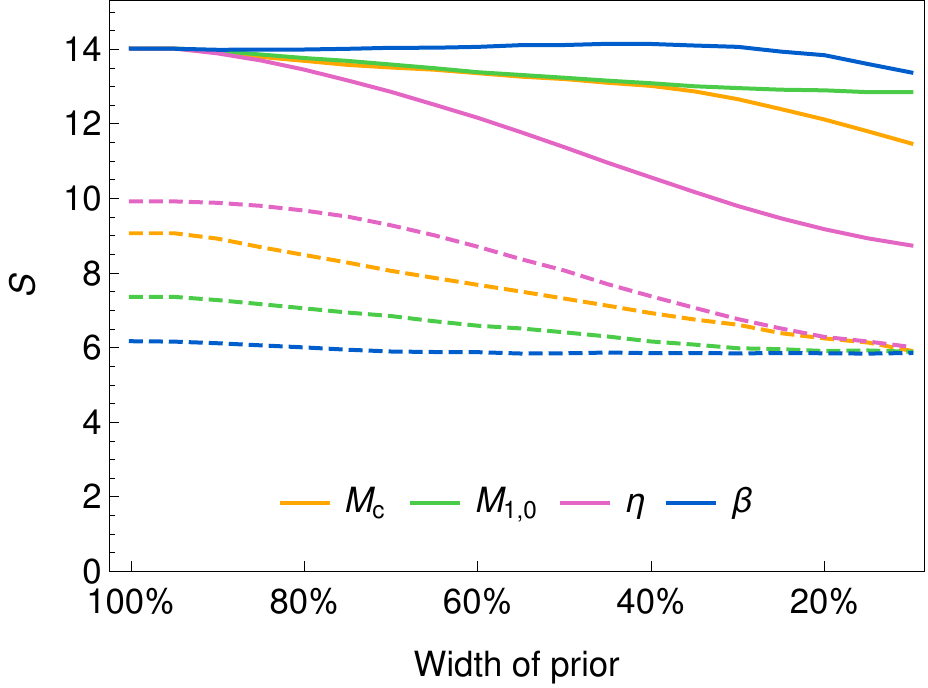}
\caption{The same as Figure~\ref{fig:prior-bounds-all}, except that only the large-scale power spectrum ($\ell<2,000$) is used.}
\label{fig:prior-bounds-ps2}
\end{figure}

The results we show in \S~\ref{sec:results-inference} assume wide priors on the baryonic parameters constrained only by WL, but we note that external observations or theoretical models can give stricter priors. Ideally, the accuracy of these parameters can be high enough so that the uncertainties of the observables caused by baryonic physics is negligible relative to those of cosmological parameters. In this section, we explore which of the baryonic parameters are most important in degrading our ability to constrain $\Om$ and $\s8$.

We refer to the full width of the prior on each baryonic parameter as 100\%, and we then gradually shrink the width of each prior linearly on a logarithmic scale until both the upper and lower bounds reach the fiducial value (we refer to the width of which as 0\%). We can manipulate the priors on the baryonic parameters in two simple ways: 1) adjust the width of prior on one parameter between 100\% and 0\% while keeping the other three priors at 100\%, or 2) adjust the width of prior on one parameter and fix the other three parameters at their fiducial values (i.e. assume 0\% priors on the latter). The resulting $1\sigma$ credible areas are shown in Figure~\ref{fig:prior-bounds-all} as a function of the prior width. They indicate the following: 
\begin{itemize}
\item If we have little knowledge about baryons and have to resort to wide priors, $\eta$ is the most important parameter, with the largest potential to improve the constraint on $\Om$ and $\s8$ (solid purple curve in Figure~\ref{fig:prior-bounds-all}).
\item If we can determine baryonic parameters accurately, $M_\mathrm{c}$ becomes the most important limiting parameter, i.e. a biased $M_\mathrm{c}$ is the least tolerable (dashed orange curve in Figure~\ref{fig:prior-bounds-all}).
\item $M_{1,0}$ and $\beta$ affect the constraint the least, regardless of our prior knowledge on the other parameters.
\end{itemize}

We also show the corresponding results in Figure~\ref{fig:prior-bounds-ps2} restricted to the large-scale only power spectrum ($\ell<2,000$). We find that in both ways of changing priors, $\eta$ is then the most important parameter, followed by $M_{1,0}$. This is consistent with the response curves in Figure~\ref{fig:parameter-response}, where $\eta$ is dominant, but only dominant at large scales.

A few observational results can be used to find the best-fitting values of the baryonic parameters. Here, we list some of these with uncertainties and their width relative to those of the priors we employ. As mentioned by \citet{schneider2015}, \citet{ade2013} suggests $\eta=0.3-0.7$ ($30\%$ of the width of our prior) from the radial profile of the cluster gas fraction; X-ray data \citep{sun2009, vikhlinin2009, gonzalez2013} suggests $\beta=0.3-0.8$ ($20\%$ width) and $M_\mathrm{c}=0.2-2.0\ee{14}h^{-1}\Msun$ ($26\%$ width) from the bounded gas fraction--$M_{500}$ relation. Stellar mass--halo mass relations from various sources listed by \citet{behroozi2013} suggests $M_{1,0}=(2-10)\ee{11}h^{-1}\Msun$ ($34\%$ width). Additionally, the matter spectrum using BCM has been fit to those obtained in hydrodynamical simulations to infer baryonic parameters by \citet{arico2020}. Interestingly, the range of simulations examined in that study generally agree on the value of $\eta$ to be $0.1-1.0$ (70\% width) but not on the other baryonic parameters.

\subsection{Born approximation}

\begin{figure*}[!ht]
\centering
\includegraphics[width=17.5cm]{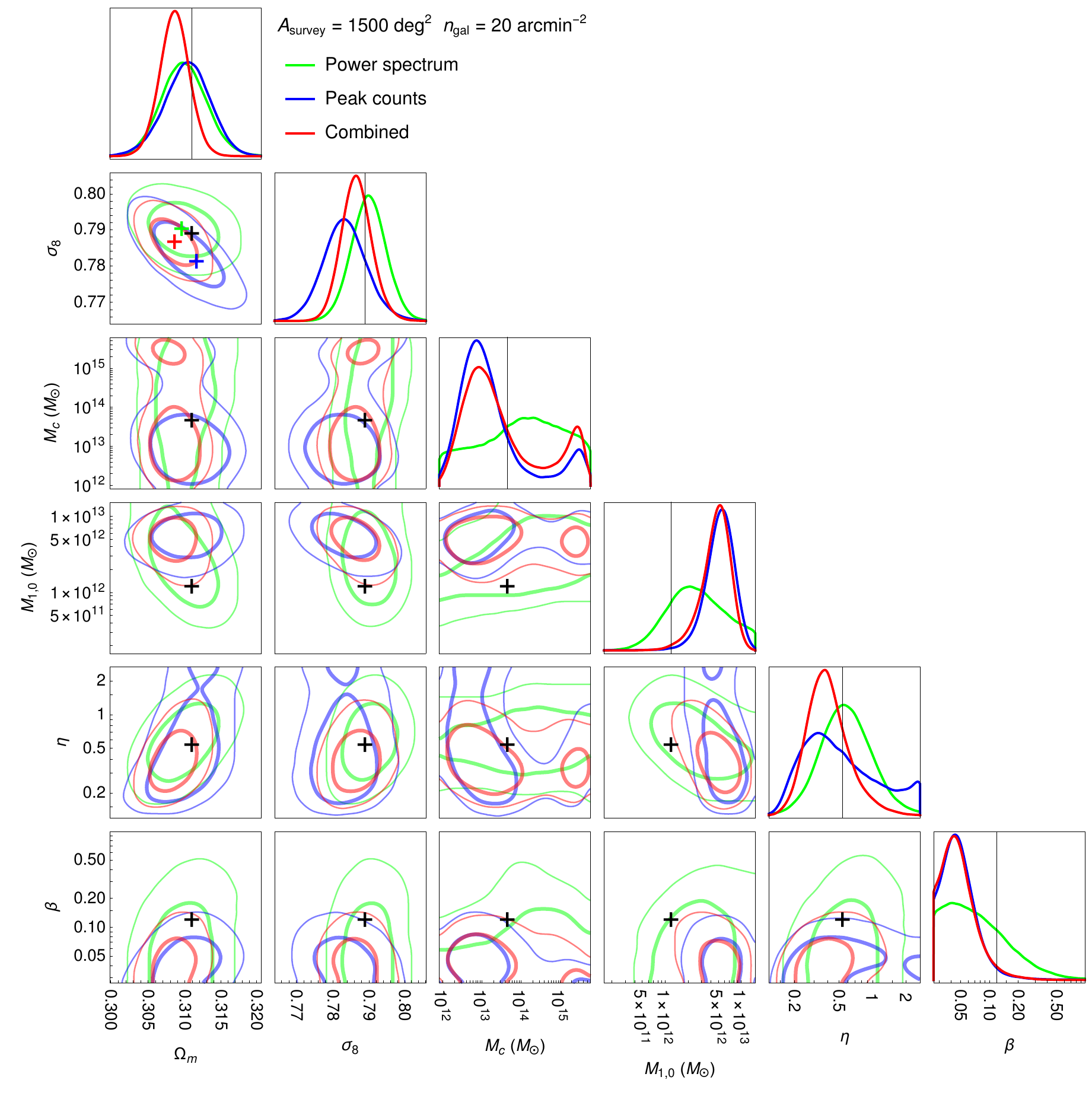}
\caption{Same as Figure~\ref{fig:cornerplot-rt-1500-20}, but the model $\mathbf{y}(\theta)$ used to fit the (fully ray-traced) mock data $\mathbf{y}_0$ employs the Born approximation. The colored crosses in the $\Om-\s8$ panel shows the maxima of the posterior distributions marginalized over the baryonic parameters. }
\label{fig:cornerplot-born-1500-20}
\end{figure*}

\begin{figure*}[!ht]
\centering
\includegraphics[width=14.5cm]{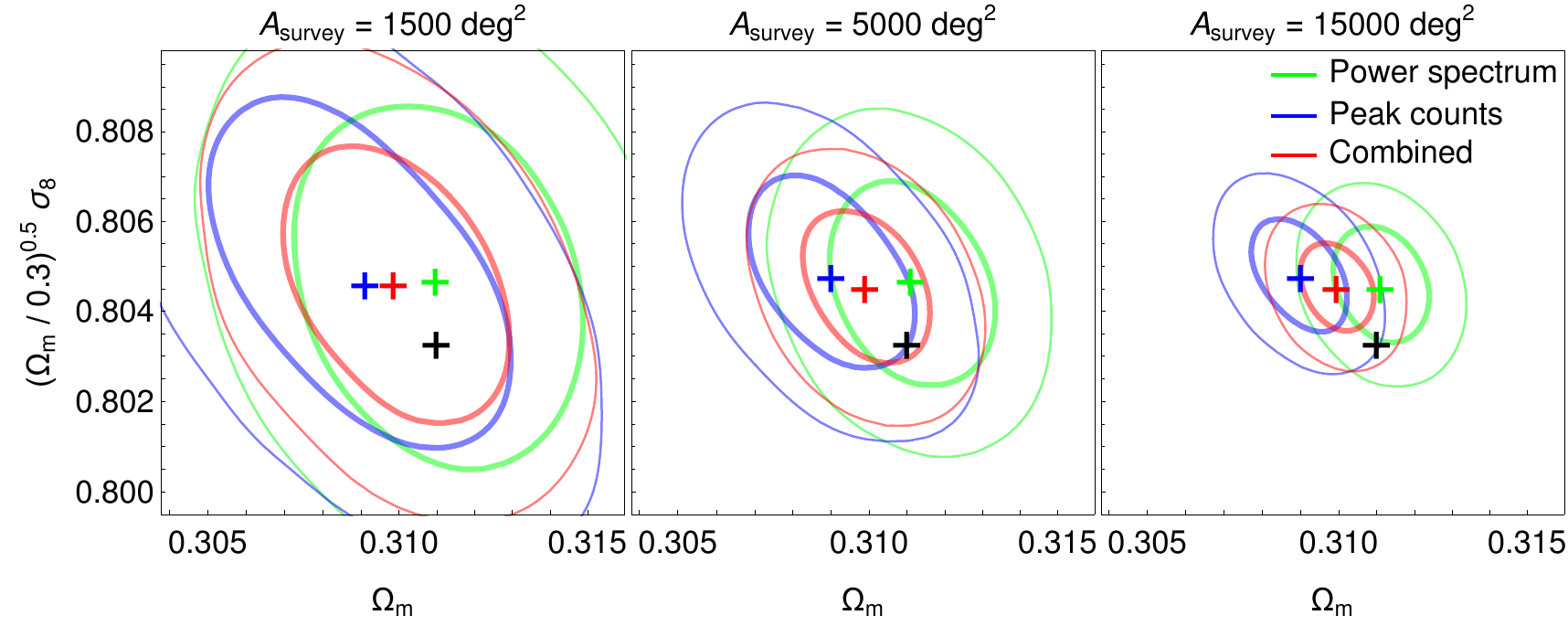}
\caption{The posterior distributions with $\mathbf{y}(\theta)$ using the Born approximation and fixing baryonic parameters to their fiducial values. The supposed survey has a galaxy density of $n_\mathrm{gal}=20\unit{arcmin^{-2}}$ and an area of $1,500$, $5,000$, or $15,000\unit{deg^2}$. The thick and thin contours show the $1\sigma$ and $2\sigma$ credible regions, respectively. The black crosses mark the fiducial cosmology, and the colored crosses mark the maximum likelihood cosmology inferred from the different statistics.}
\label{fig:born-bias-fiducial}
\end{figure*}

The large majority of previous works investigating baryonic effects obtained lensing power spectra via a direct projection of the three-dimensional matter power spectrum \citep{rudd2008, zentner2008, mead2015}, or used the Born approximation in place of full ray-tracing \citep[][but see \citealt{osato2021} for a recent exception]{semboloni2011, fong2019, huang2019, weiss2019}.

In the Born approximation, the convergence at position $\boldsymbol{\theta}$ from source comoving distance $\chi_\mathrm{s}$ is approximated by taking the lowest order term of the ray-tracing calculations: 
\begin{equation}
\kappa_\mathrm{born}(\boldsymbol{\theta})=\frac{3H_0^2\Om}{2c^2}\int_0^{\chi_\mathrm{s}}\frac{\chi\,\mathrm{d}\chi_\mathrm{s}}{a(\chi)}W(\chi,\chi_\mathrm{s})\delta(\boldsymbol{\theta},\chi),
\end{equation}
where $a$ denotes the scale factor, $W(\chi,\chi_\mathrm{s})=1-\chi/\chi_\mathrm{s}$ the lensing kernel, and $\delta$ the density contrast. Note that the calculation of $\kappa_\mathrm{born}$ only depends on the density contrast values along the line of sight, which itself is approximated as a straight line. It makes the Born approximation much cheaper than ray-tracing in terms of computational cost. But assuming that the observed lensing effects can only be resembled by ray-tracing, we need to assess the biases introduced by the Born approximation whenever applied to WL surveys.

We build a set of convergence maps using the Born approximation in parallel with the ray-tracing maps, and we apply them to two scenarios of parameter inference. First, we replace the model $\mathbf{y}(\theta)$ in Equation~\ref{eqn: delta-y} with the model derived from the Born approximation convergence maps (but not replacing $\mathbf{y}_0$), and infer cosmological and baryonic parameters, the sampled posteriors of which is shown in Figure~\ref{fig:cornerplot-born-1500-20}. We find that the Born approximation can cause a large bias in this scenario for an HSC-like survey: the power spectrum method is only slightly biased ($<0.5\sigma$ for $\Om$ and $\s8$), but peak counts have a $1\sigma$ bias in the two cosmological parameters and $\sim2\sigma$ biases on $M_{1,0}$ and $\beta$. Second, we fix the baryonic parameters to their fiducial values and infer $\Om$ and $\s8$. As is shown in Figure~\ref{fig:born-bias-fiducial}, cosmological parameters are biased by a similar absolute amount, but since the constraints are tighter with larger survey area, the biases are more statistically significant. For the power spectrum, the significance reaches $1\sigma$ when $A_\mathrm{survey}=15,000\unit{deg^2}$, and for peak counts, the significance reaches $1\sigma$ when $A_\mathrm{survey}=5,000\unit{deg^2}$. Similar results have also been noted by \citet{petri2017} for the power spectrum and moments of the convergence field (in the absence of baryons). 

In summary, using the Born approximation instead of ray-tracing has a greater impact on the peak counts than on the power spectrum, possibly because peak counts rely more heavily on small-scale information where there are larger the deviations between the Born approximation and ray-tracing. In an HSC-like survey, ray-tracing is necessary for inferring baryonic parameters or to correctly marginalize over baryons but only marginally necessary for inferring $\Om$ and $\s8$ for fixed baryonic parameters.

\subsection{The dependence of covariances on baryonic parameters}
\label{sec:cov-dependence}

\begin{figure}[!ht]
\centering
\includegraphics[width=7.5cm]{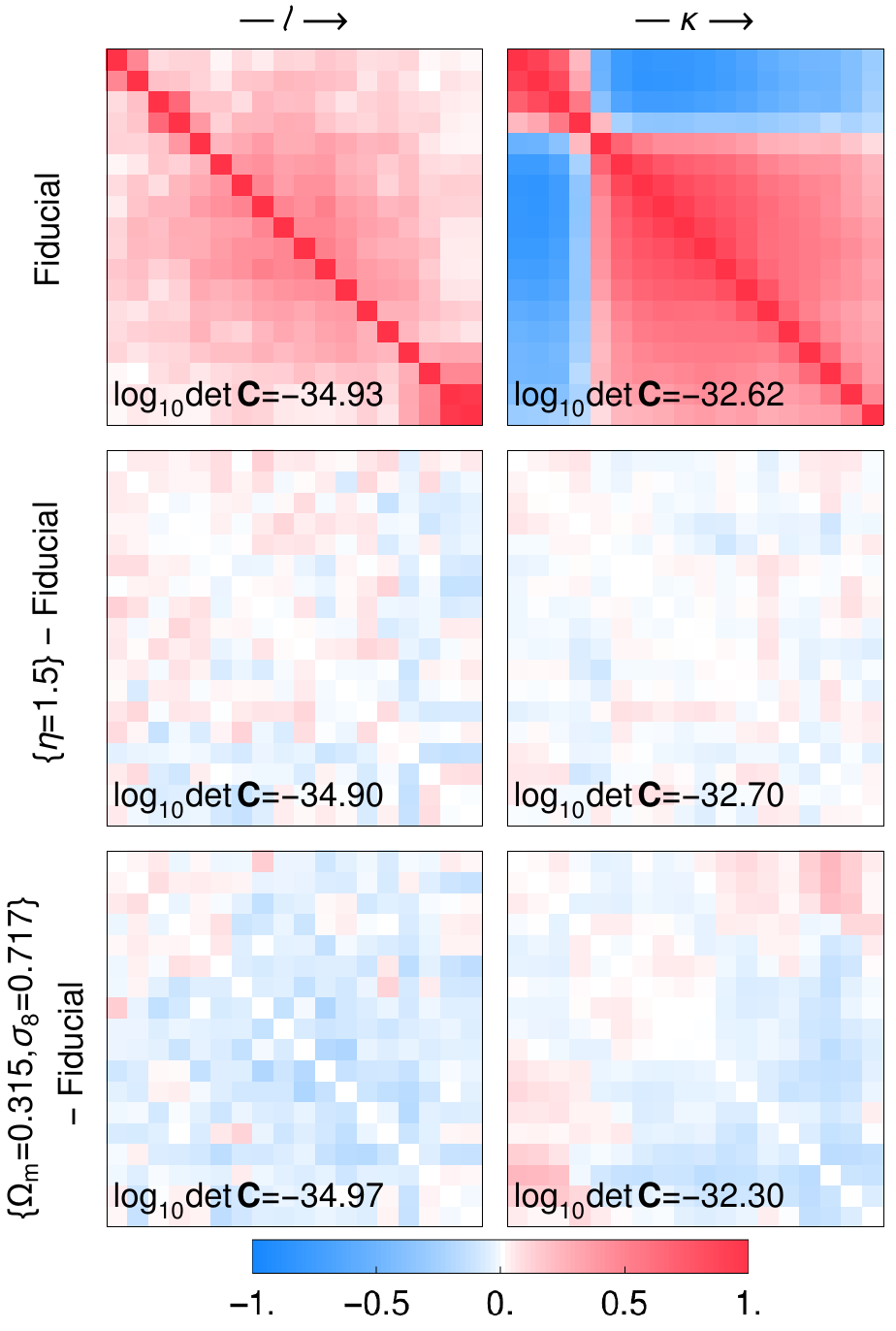}
\caption{The correlation matrix $\rho$ of the observables in the fiducial model (upper panels), the difference of $\rho$ between the $\eta=0.54$ model (all other parameters remain unchanged) and the fiducial model (middle panels), and the difference between the ($\Om=0.315,\s8=0.717$) model and the fiducial model. The determinants of the covariance matrices in all three models are shown at the bottom left corner of each panel.}
\label{fig:cov-dependence}
\end{figure}

In Figure~\ref{fig:cov-dependence}, we show how covariances change with $\eta$, the most important baryonic parameter. As an indication for the importance of the $\eta$-dependence, we compare the determinants of the covariance matrices and the correlations between the observables with two $\eta$ values---0.54 (fiducial value) and 1.5. We find that there are only slight differences between them: the determinant changes by only 7\% and 20\% for the two statistics. By comparison, the determinant varies by 10\% and a factor of $\approx2$, respectively, due to a 10\% change in $\s8$ ($0.789\to0.717$), which is a typical level of uncertainty from previous weak lensing surveys \citep[see e.g.][]{hikage2019}.

\section{Conclusions}
\label{sec:conclusions}

In this work, we adopted a start-of-the-art baryonic correction model (BCM) to investigate the impact of baryons on the inference of cosmological parameters $\Om$ and $\s8$ from Gaussian and non-Gaussian WL statistics (the power spectrum and peak counts). In order to be able to incorporate the baryonic effects into a large suite of ray-tracing simulations, we employed a low computational-cost methodology where we ``painted'' baryons onto $N$-body simulations by replacing the particles in DM haloes by images of analytical profiles that contain baryons. 

We summarize our main findings as follows:

\begin{enumerate}
\item The difference in column densities between DM haloes from the $N$-body simulations and their NFW-halo replacements is mainly the quadrupole component, which explains most of the discrepancy in the column density power spectrum at $k\gtrsim3\,h\unit{Mpc^{-1}}$.
\item If the baryonic parameters are inferred simultaneously with $\Omega_m$ and $\sigma_8$ and being marginalized over, the area of the $1\sigma$ region in $\Om-\s8$ space will be $2-4$ times larger than that with baryonic parameters fixed. This factor is larger for peak counts and for higher galaxy densities (less noisy data). 
\item The quality of constraints on $\Om$ and $\s8$ in terms of the $1\sigma$ area improves the most when $\eta$ and $M_\mathrm{c}$ have tighter priors, while $M_{1,0}$ and $\beta$ barely affects the $1\sigma$ area.
\item In an HSC-like survey, expanding the range of lensing power spectrum from $\ell<2,000$ to $\ell<12,000$ can improve the constraints on $\Om$ and $\s8$ by 11\% when marginalized over baryonic parameters and by 24\% when baryonic parameters are fixed. These improvements will be larger at higher galaxy densities.
\item Baryonic physics, with the BCM models we adopted, and in the absence of strong priors, affects the power spectrum on both small and large scales; a limit of $\ell<2,000$ still requires the modelling of baryons.
\item Using the Born approximation to generate convergence maps has a greater impact on peak counts than power spectrum, but is necessary to correctly model baryonic effects and marginalize over them, for both statistics in an HSC-like survey.
\end{enumerate}

We note two caveats in this work that do not affect our finding above but can be a problem if we apply the same methods to observational data in the future. We need to carry the shape information of the DM haloes from the $N$-body simulation to the analytical profiles to remove the halo replacement residuals. We also need to generate more realizations of the convergence map for each combination of parameters so that when comparing to the real data, the statistics are not biased due to insufficient randomization. More generally, before they can be applied to real data, a better calibration of the specific parameterized BCM will be necessary against physical hydrodynamical models, including a range of baryon physics and their impact on peak counts.

\section*{Acknowledgements}

We thank Colin Hill and Lam Hui for useful discussions and acknowledge support by NASA ATP grant 80NSSC18K1093. We acknowledge support from the NSF XSEDE facility for data analysis in this study.

\bibliography{main}{}

\begin{thebibliography}{}
\expandafter\ifx\csname natexlab\endcsname\relax\def\natexlab#1{#1}\fi
\providecommand{\url}[1]{\href{#1}{#1}}
\providecommand{\dodoi}[1]{doi:~\href{http://doi.org/#1}{\nolinkurl{#1}}}
\providecommand{\doeprint}[1]{\href{http://ascl.net/#1}{\nolinkurl{http://ascl.net/#1}}}
\providecommand{\doarXiv}[1]{\href{https://arxiv.org/abs/#1}{\nolinkurl{https://arxiv.org/abs/#1}}}

\bibitem[{Abbott {et~al.}(2005)Abbott, Aldering, Annis, {et~al.}}]{des2005}
Abbott, T., Aldering, G., Annis, J., {et~al.} 2005, arXiv preprint
  astro-ph/0510346

\bibitem[{Ade {et~al.}(2013)Ade, Aghanim, Arnaud, Ashdown, Atrio-Barandela,
  Aumont, Baccigalupi, Balbi, Banday, Barreiro, {et~al.}}]{ade2013}
Ade, P.~A., Aghanim, N., Arnaud, M., {et~al.} 2013, Astronomy \& Astrophysics,
  550, A131

\bibitem[{Aihara {et~al.}(2018)Aihara, Arimoto, Armstrong, Arnouts, Bahcall,
  Bickerton, Bosch, Bundy, Capak, Chan, {et~al.}}]{aihara2018}
Aihara, H., Arimoto, N., Armstrong, R., {et~al.} 2018, Publications of the
  Astronomical Society of Japan, 70, S4

\bibitem[{Aric{\`o} {et~al.}(2020)Aric{\`o}, Angulo, Hern{\'a}ndez-Monteagudo,
  Contreras, Zennaro, Pellejero-Iba{\~n}ez, \& Rosas-Guevara}]{arico2020}
Aric{\`o}, G., Angulo, R.~E., Hern{\'a}ndez-Monteagudo, C., {et~al.} 2020,
  Monthly Notices of the Royal Astronomical Society, 495, 4800

\bibitem[{Bartelmann \& Schneider(2001)}]{bartelmann2001}
Bartelmann, M., \& Schneider, P. 2001, Physics Reports, 340, 291

\bibitem[{Behroozi {et~al.}(2013)Behroozi, Wechsler, \& Conroy}]{behroozi2013}
Behroozi, P.~S., Wechsler, R.~H., \& Conroy, C. 2013, The Astrophysical
  Journal, 770, 57

\bibitem[{Behroozi {et~al.}(2012)Behroozi, Wechsler, \& Wu}]{behroozi2012}
Behroozi, P.~S., Wechsler, R.~H., \& Wu, H.-Y. 2012, The Astrophysical Journal,
  762, 109

\bibitem[{de~Jong {et~al.}(2013)de~Jong, Kleijn, Kuijken, Valentijn,
  {et~al.}}]{dejong2013}
de~Jong, J.~T., Kleijn, G. A.~V., Kuijken, K.~H., Valentijn, E.~A., {et~al.}
  2013, Experimental Astronomy, 35, 25

\bibitem[{Fang \& Haiman(2007)}]{fang2007}
Fang, W., \& Haiman, Z. 2007, Physical Review D, 75, 043010

\bibitem[{Fong {et~al.}(2019)Fong, Choi, Catlett, Lee, Peel, Bowyer, King, \&
  McCarthy}]{fong2019}
Fong, M., Choi, M., Catlett, V., {et~al.} 2019, Monthly Notices of the Royal
  Astronomical Society, 488, 3340

\bibitem[{Gonzalez {et~al.}(2013)Gonzalez, Sivanandam, Zabludoff, \&
  Zaritsky}]{gonzalez2013}
Gonzalez, A.~H., Sivanandam, S., Zabludoff, A.~I., \& Zaritsky, D. 2013, The
  Astrophysical Journal, 778, 14

\bibitem[{Hamana {et~al.}(2020)Hamana, Shirasaki, Miyazaki, Hikage, Oguri,
  More, Armstrong, Leauthaud, Mandelbaum, Miyatake, {et~al.}}]{hamana2020}
Hamana, T., Shirasaki, M., Miyazaki, S., {et~al.} 2020, Publications of the
  Astronomical Society of Japan, 72, 16

\bibitem[{Hartlap {et~al.}(2007)Hartlap, Simon, \& Schneider}]{hartlap2007}
Hartlap, J., Simon, P., \& Schneider, P. 2007, Astronomy \& Astrophysics, 464,
  399

\bibitem[{Heymans {et~al.}(2012)Heymans, Van~Waerbeke, Miller, Erben,
  Hildebrandt, Hoekstra, Kitching, Mellier, Simon, Bonnett,
  {et~al.}}]{heymans2012}
Heymans, C., Van~Waerbeke, L., Miller, L., {et~al.} 2012, Monthly Notices of
  the Royal Astronomical Society, 427, 146

\bibitem[{Hikage {et~al.}(2019)Hikage, Oguri, Hamana, More, Mandelbaum, Takada,
  K{\"o}hlinger, Miyatake, Nishizawa, Aihara, {et~al.}}]{hikage2019}
Hikage, C., Oguri, M., Hamana, T., {et~al.} 2019, Publications of the
  Astronomical Society of Japan, 71, 43

\bibitem[{Hilbert {et~al.}(2009)Hilbert, Hartlap, White, \&
  Schneider}]{hilbert2009}
Hilbert, S., Hartlap, J., White, S., \& Schneider, P. 2009, Astronomy \&
  Astrophysics, 499, 31

\bibitem[{Hoekstra \& Jain(2008)}]{hoekstra2008}
Hoekstra, H., \& Jain, B. 2008, Annual Review of Nuclear and Particle Science,
  58, 99

\bibitem[{Huang {et~al.}(2019)Huang, Eifler, Mandelbaum, \&
  Dodelson}]{huang2019}
Huang, H.-J., Eifler, T., Mandelbaum, R., \& Dodelson, S. 2019, Monthly Notices
  of the Royal Astronomical Society, 488, 1652

\bibitem[{Ivezi{\'c} {et~al.}(2019)Ivezi{\'c}, Kahn, Tyson, Abel, Acosta,
  Allsman, Alonso, AlSayyad, Anderson, Andrew, {et~al.}}]{ivezic2019}
Ivezi{\'c}, {\v{Z}}., Kahn, S.~M., Tyson, J.~A., {et~al.} 2019, The
  Astrophysical Journal, 873, 111

\bibitem[{Jain {et~al.}(2000)Jain, Seljak, \& White}]{jain2000}
Jain, B., Seljak, U., \& White, S. 2000, The Astrophysical Journal, 530, 547

\bibitem[{Jenkins {et~al.}(2001)Jenkins, Frenk, White, Colberg, Cole, Evrard,
  Couchman, \& Yoshida}]{jenkins2001}
Jenkins, A., Frenk, C., White, S.~D., {et~al.} 2001, Monthly Notices of the
  Royal Astronomical Society, 321, 372

\bibitem[{Jing {et~al.}(2006)Jing, Zhang, Lin, Gao, \& Springel}]{jing2006}
Jing, Y., Zhang, P., Lin, W., Gao, L., \& Springel, V. 2006, The Astrophysical
  Journal Letters, 640, L119

\bibitem[{Joudaki {et~al.}(2016)Joudaki, Blake, Heymans, Choi, Harnois-Deraps,
  Hildebrandt, Joachimi, Johnson, Mead, Parkinson, {et~al.}}]{joudaki2016}
Joudaki, S., Blake, C., Heymans, C., {et~al.} 2016, Monthly Notices of the
  Royal Astronomical Society, stw2665

\bibitem[{{Kilbinger}(2015)}]{kilbinger2015}
{Kilbinger}, M. 2015, Reports on Progress in Physics, 78, 086901,
  \dodoi{10.1088/0034-4885/78/8/086901}

\bibitem[{K{\"o}hlinger {et~al.}(2016)K{\"o}hlinger, Viola, Valkenburg,
  Joachimi, Hoekstra, \& Kuijken}]{kohlinger2016}
K{\"o}hlinger, F., Viola, M., Valkenburg, W., {et~al.} 2016, Monthly Notices of
  the Royal Astronomical Society, 456, 1508

\bibitem[{K{\"o}hlinger {et~al.}(2017)K{\"o}hlinger, Viola, Joachimi, Hoekstra,
  Van~Uitert, Hildebrandt, Choi, Erben, Heymans, Joudaki,
  {et~al.}}]{kohlinger2017}
K{\"o}hlinger, F., Viola, M., Joachimi, B., {et~al.} 2017, Monthly Notices of
  the Royal Astronomical Society, 471, 4412

\bibitem[{Laureijs {et~al.}(2011)Laureijs, Amiaux, Arduini, Augueres,
  Brinchmann, Cole, Cropper, Dabin, Duvet, Ealet, {et~al.}}]{laureijs2011}
Laureijs, R., Amiaux, J., Arduini, S., {et~al.} 2011, arXiv preprint
  arXiv:1110.3193

\bibitem[{Lewis {et~al.}(2000)Lewis, Challinor, \& Lasenby}]{lewis2000}
Lewis, A., Challinor, A., \& Lasenby, A. 2000, The Astrophysical Journal, 538,
  473

\bibitem[{Mead {et~al.}(2015)Mead, Peacock, Heymans, Joudaki, \&
  Heavens}]{mead2015}
Mead, A., Peacock, J., Heymans, C., Joudaki, S., \& Heavens, A. 2015, Monthly
  Notices of the Royal Astronomical Society, 454, 1958

\bibitem[{Mohammed {et~al.}(2014)Mohammed, Martizzi, Teyssier, \&
  Amara}]{mohammed2014}
Mohammed, I., Martizzi, D., Teyssier, R., \& Amara, A. 2014, arXiv preprint
  arXiv:1410.6826

\bibitem[{Nelson {et~al.}(2019)Nelson, Springel, Pillepich, Rodriguez-Gomez,
  Torrey, Genel, Vogelsberger, Pakmor, Marinacci, Weinberger,
  {et~al.}}]{nelson2019}
Nelson, D., Springel, V., Pillepich, A., {et~al.} 2019, Computational
  Astrophysics and Cosmology, 6, 1

\bibitem[{Osato {et~al.}(2021)Osato, Liu, \& Haiman}]{osato2021}
Osato, K., Liu, J., \& Haiman, Z. 2021, Monthly Notices of the Royal
  Astronomical Society, 502, 5593

\bibitem[{Petri(2016)}]{petri2016b}
Petri, A. 2016, Astronomy and Computing, 17, 73

\bibitem[{{Petri} {et~al.}(2016){Petri}, {Haiman}, \& {May}}]{petri2016a}
{Petri}, A., {Haiman}, Z., \& {May}, M. 2016, \prd, 93, 063524,
  \dodoi{10.1103/PhysRevD.93.063524}

\bibitem[{Petri {et~al.}(2017)Petri, Haiman, \& May}]{petri2017}
Petri, A., Haiman, Z., \& May, M. 2017, Physical Review D, 95, 123503

\bibitem[{Refregier(2003)}]{refregier2003b}
Refregier, A. 2003, Annual Review of Astronomy and Astrophysics, 41, 645

\bibitem[{Rudd {et~al.}(2008)Rudd, Zentner, \& Kravtsov}]{rudd2008}
Rudd, D.~H., Zentner, A.~R., \& Kravtsov, A.~V. 2008, The Astrophysical
  Journal, 672, 19

\bibitem[{Schaye {et~al.}(2010)Schaye, Vecchia, Booth, Wiersma, Theuns, Haas,
  Bertone, Duffy, McCarthy, \& van~de Voort}]{schaye2010}
Schaye, J., Vecchia, C.~D., Booth, C., {et~al.} 2010, Monthly Notices of the
  Royal Astronomical Society, 402, 1536

\bibitem[{Schneider \& Teyssier(2015)}]{schneider2015}
Schneider, A., \& Teyssier, R. 2015, Journal of Cosmology and Astroparticle
  Physics, 2015, 049

\bibitem[{Schneider {et~al.}(2019)Schneider, Teyssier, Stadel, Chisari,
  Le~Brun, Amara, \& Refregier}]{schneider2019}
Schneider, A., Teyssier, R., Stadel, J., {et~al.} 2019, Journal of Cosmology
  and Astroparticle Physics, 2019, 020

\bibitem[{Schneider \& Bridle(2010)}]{schneider2010}
Schneider, M.~D., \& Bridle, S. 2010, Monthly Notices of the Royal Astronomical
  Society, 402, 2127

\bibitem[{Semboloni {et~al.}(2011)Semboloni, Hoekstra, Schaye, van Daalen, \&
  McCarthy}]{semboloni2011}
Semboloni, E., Hoekstra, H., Schaye, J., van Daalen, M.~P., \& McCarthy, I.~G.
  2011, Monthly Notices of the Royal Astronomical Society, 417, 2020

\bibitem[{Sif{\'o}n {et~al.}(2015)Sif{\'o}n, Hoekstra, Cacciato, Viola,
  K{\"o}hlinger, van~der Burg, Sand, \& Graham}]{sifon2015}
Sif{\'o}n, C., Hoekstra, H., Cacciato, M., {et~al.} 2015, Astronomy \&
  Astrophysics, 575, A48

\bibitem[{Sobol'(1967)}]{sobol1967}
Sobol', I.~M. 1967, Zhurnal Vychislitel'noi Matematiki i Matematicheskoi
  Fiziki, 7, 784

\bibitem[{Spergel {et~al.}(2015)Spergel, Gehrels, Baltay, Bennett,
  Breckinridge, Donahue, Dressler, Gaudi, Greene, Guyon,
  {et~al.}}]{spergel2015}
Spergel, D., Gehrels, N., Baltay, C., {et~al.} 2015, arXiv preprint
  arXiv:1503.03757

\bibitem[{Springel(2005)}]{springel2005}
Springel, V. 2005, Monthly notices of the royal astronomical society, 364, 1105

\bibitem[{Sun {et~al.}(2009)Sun, Voit, Donahue, Jones, Forman, \&
  Vikhlinin}]{sun2009}
Sun, M., Voit, G., Donahue, M., {et~al.} 2009, The Astrophysical Journal, 693,
  1142

\bibitem[{Vikhlinin {et~al.}(2009)Vikhlinin, Burenin, Ebeling, Forman,
  Hornstrup, Jones, Kravtsov, Murray, Nagai, Quintana,
  {et~al.}}]{vikhlinin2009}
Vikhlinin, A., Burenin, R., Ebeling, H., {et~al.} 2009, The Astrophysical
  Journal, 692, 1033

\bibitem[{Weiss {et~al.}(2019)Weiss, Schneider, Sgier, Kacprzak, Amara, \&
  Refregier}]{weiss2019}
Weiss, A.~J., Schneider, A., Sgier, R., {et~al.} 2019, Journal of Cosmology and
  Astroparticle Physics, 2019, 011

\bibitem[{White(2004)}]{white2004}
White, M. 2004, Astroparticle Physics, 22, 211

\bibitem[{Yang {et~al.}(2013)Yang, Kratochvil, Huffenberger, Haiman, \&
  May}]{yang2013}
Yang, X., Kratochvil, J.~M., Huffenberger, K., Haiman, Z., \& May, M. 2013,
  Physical Review D, 87, 023511

\bibitem[{Zentner {et~al.}(2008)Zentner, Rudd, \& Hu}]{zentner2008}
Zentner, A.~R., Rudd, D.~H., \& Hu, W. 2008, Physical Review D, 77, 043507

\bibitem[{Zhan \& Knox(2004)}]{zhan2004}
Zhan, H., \& Knox, L. 2004, The Astrophysical Journal Letters, 616, L75

\end{thebibliography}
\bibliographystyle{aasjournal}

\end{document}